\documentclass[prb,preprint]{revtex4} 

\newcommand{\beq}{\begin{equation}}
\newcommand{\eeq}{\end{equation}}
\newcommand{\beqarray}{\begin{eqnarray}}
\newcommand{\eeqarray}{\end{eqnarray}}
\newcommand{\bx}{\bm{x}}
\newcommand{\bA}{\bm{A}}

\newcommand{\del}{\bm{\nabla}}
\newcommand{\bk}{\bm{k}}

\usepackage{graphicx}
\usepackage{amssymb}
\usepackage{bm}
\usepackage{amsmath}

\begin{document}

\title{On an identity for the volume integral of the square of a vector
field. Remark on a paper by A. M. Stewart}

\author{Loyal Durand} \email{ldurand@hep.wisc.edu}
\affiliation{Department of Physics, University of Wisconsin-Madison,
Madison, Wisconsin 53706}

\maketitle

Gubarev, Stodolsky, and Zakharov\cite{Gubarev} have noted the following
identity for vector fields that vanish sufficiently rapidly at spatial
infinity,
\beqarray
\label{identity}
\int\!d^3x\bA(\bx)\cdot\bA(\bx
&=& \frac{1}{4\pi}\!\int\!d^3x\,d^3x'\frac{1}{|\bx-\bx'|}\Big[
\big(\del\cdot\bA(\bx)\big)\big(\del'\cdot\bA(\bx')\big)  \nonumber
\\
&&{}
+ \big(\del\times\bA(\bx)\big)\cdot
\big(\del'\times\bA(\bx')\big)\Big], 
\eeqarray
and used it to investigate properties of the vector potential in quantum field
theory. Stewart\cite{Stewart} has shown that the identity is also of general
interest in classical electromagnetic theory. It can be used, for example,
to derive easily interpreted expressions for the energies in time-dependent
electric and magnetic fields. 

The existence of this identity is not obvious. In
Ref.~\onlinecite{Gubarev} it is not proven, but it is noted  that
it follows from the momentum-space relation
$(\bk\times\tilde{\bA})^2=\bk^2\tilde{\bA}^2-(\bk\cdot\tilde{\bA})^2$
where $\tilde{\bA}(\bk)$ is the Fourier transform of the vector field. The position-space derivation given by Stewart is based
on the Helmholtz decomposition of a three-dimensional vector field into
irrotational and solenoidal parts.  We give here an alternative position-space 
derivation that uses
only familiar operations starting from the relations
\begin{equation}
\label{deltafunction}
\delta^3(\bx-\bx')=-\frac{1}{4\pi}\nabla^2\frac{1}{|\bx-\bx'|)} 
\end{equation}
for the Dirac delta function, and
\begin{equation}
\label{doublecurl}
\del\times(\del\times \bA)=\del(\del\cdot\bA)-\nabla^2\!\bA.
\end{equation}
Both are  familiar to students and are used
in the solution of Poisson's equation for the potential of a point charge
and the derivation of the wave equation from Maxwell's equations.  

By using Eq.~(\ref{deltafunction}), we obtain
\begin{align}
\label{identity1}
\int\!d^3x\bA(\bx)\cdot\bA(\bx)&=\!\int
d^3x\,d^3x'\bA(\bx)\cdot\delta^3(\bx-\bx')\bA(\bx') \nonumber \\
&= -\frac{1}{4\pi}\!\int\!d^3x\,d^3x'\bA(\bx) \cdot\Big(
\nabla^{'2}\frac{1}{|\bx-\bx'|}\Big)\bA(\bx').
\end{align}
If we integrate by parts over $\bx'$ and use the assumed rapid vanishing
of
$\bA(\bx')$ for $|\bx'|\rightarrow\infty$, we can transfer the action of the
derivatives from the factor $1/|\bx-\bx'|$ to $\bA(\bx')$ without acquiring
extra surface terms. We use the identity (\ref{doublecurl}) for the
double curl, make another partial integration, and  rewrite the
result as
\begin{subequations}
\begin{align}
\label{identity2}
\int\!d^3x \bA(\bx)\!\cdot\!\bA(\bx)
&= -\frac{1}{4\pi}\!\int\!d^3x\,d^3x'\bA(\bx)
\cdot\frac{1}{|\bx-\bx'|}\nabla^{'2}\bA(\bx')  \\
&= -\frac{1}{4\pi}\!\int\!d^3x\,d^3x'\bA(\bx)
\cdot\frac{1}{|\bx-\bx'|}\Big[\del'\big(\del'\cdot\bA(\bx')\big) -
\del'\times\big(\del'\times \bA(\bx')\big)\Big] \nonumber \\
&= \frac{1}{4\pi}\!\int\!d^3x\,d^3x'\Big[\Big(\bA(\bx)
\cdot\del'\frac{1}{|\bx-\bx'|}\Big)(\del'\cdot\bA(\bx') \nonumber \\
& \quad -
\bA(\bx)\cdot\Big(\del'\frac{1}{|\bx-\bx'|}\Big)
\times \big(\del'\times\bA(\bx')\big)\Big]. \label{last}
\end{align}
\end{subequations}
Now $\del'(1/|\bx-\bx'|)=-\del(1/|\bx-\bx'|)$, so with
some rearrangement of the vector products, Eq.~\eqref{last} becomes
\beqarray
\label{identity3}
\int\!d^3x\bA(\bx)\cdot\bA(\bx)
&=& -\frac{1}{4\pi}\!\int\!d^3x\,d^3x' \Big[
\big(\bA(\bx)\cdot\del\frac{1}{|\bx-\bx'|}\big)\big(\del'\cdot\bA(\bx')
\big)  \nonumber \\ &&
\quad +\big(\del\frac{1}{|\bx-\bx'|}\big)
\times\bA(\bx)\cdot\big(\del'\times\bA(\bx')\big)\Big]. 
\eeqarray
A final partial integration with respect to $\bx$  gives the desired
identity.   

The derivation generalizes easily to give an identity of the same form for the volume integral of the product of two different rapidly decreasing vector fields.


\begin{thebibliography}{99}

\bibitem{Gubarev} F. V. Gubarev, L. Stodolsky, and V. I. Zakarov, ``On the
significance of the vector potential squared," Phys. Rev. Lett. 86 (11),
2220--2222 (2001).

\bibitem{Stewart} A. M. Stewart, ``On an identity for the volume integral of
the square of a vector field," Am. J. Phys. (preceding paper).

\end{thebibliography}
\end{document}